\documentclass[aps,pre,twocolumn,letter]{revtex4}

\usepackage{amssymb}
\usepackage{amsopn}
\usepackage{graphics}
\usepackage{graphicx}
\usepackage{textcomp}
\usepackage{amsmath}
\usepackage{amsthm}
\usepackage{gensymb} 
\usepackage{bm,subfigure}
\usepackage{color}

\newcommand{\bb}{\begin{eqnarray}}
\newcommand{\ee}{\end{eqnarray}}

\begin{document}
\title{Unfolding protein with an atomic force microscope: \\ Force-fluctuation induced non-exponential kinetics }

\author{Maxime Clusel}	
\email{maxime.clusel@univ-montp2.fr}
\affiliation{Institut Laue-Langevin, 6 rue Jules Horowitz, BP156X, 38042 Grenoble cedex, France.}
\affiliation{Laboratoire Charles Coulomb UMR 5221, CNRS and Universit\'e Montpellier 2, F-34095, Montpellier, France. }
\author{Eric I. Corwin}
\email{ecorwin@uoregon.edu}
\affiliation{Department of Physics, University of Oregon, Eugene, Oregon 97403, USA.}

\date{\today}

\begin{abstract}
We show that in experimental atomic force microscopy studies of the lifetime distribution of mechanically stressed folded proteins the effects of externally applied fluctuations can not be distinguished from those of internally present fluctuations.  In certain circumstances this leads to artificially non-exponential lifetime distributions which can be misinterpreted as a signature of protein complexity.  This work highlights the importance of fully characterizing and controlling external sources of fluctuation in mechanical studies of proteins before drawing conclusions on the physics at play on the molecular level.
\end{abstract}

\maketitle

\section{Introduction}

Atomic force microscopy (AFM) has emerged as a powerful tool with which to study biologically relevant systems \cite{alonso2003,alessandrini2005}. Probing the response of a protein to an applied mechanical force allows for the direct investigation of the physical properties of the protein \cite{zlatanova2006,cohen2008}. Recent experiments have used lifetime distributions of a single folded protein under constant applied mechanical force to probe physics at the molecular scale \cite{oberhauser2001, schlierf2004}. To do so, a folded protein is put under tension and the time to unfolding is measured.  Many such individual experiments are collected (often over the course of days or even weeks) and their lifetimes combined to estimate the lifetime distribution.  Such lifetime distributions have been studied experimentally for both Ubiquitin and the 27th domain of immunoglobulin (I27) \cite{garcia2007,garcia2009}. In some instances non-exponential behavior has been reported in measurements of Ubiquitin \cite{brujic2006,brujic2007}, and associated with glassy dynamics \cite{brujic2006}. Static configurational disorder of the folded protein structure combined with the Bell model \cite{bell1978} has recently been proposed as a plausible explanation for the observed stretched-exponential unfolding time distributions \cite{raible2006,kuo2010}.\\
\begin{figure}[tb]
\includegraphics[width=.97 \columnwidth]{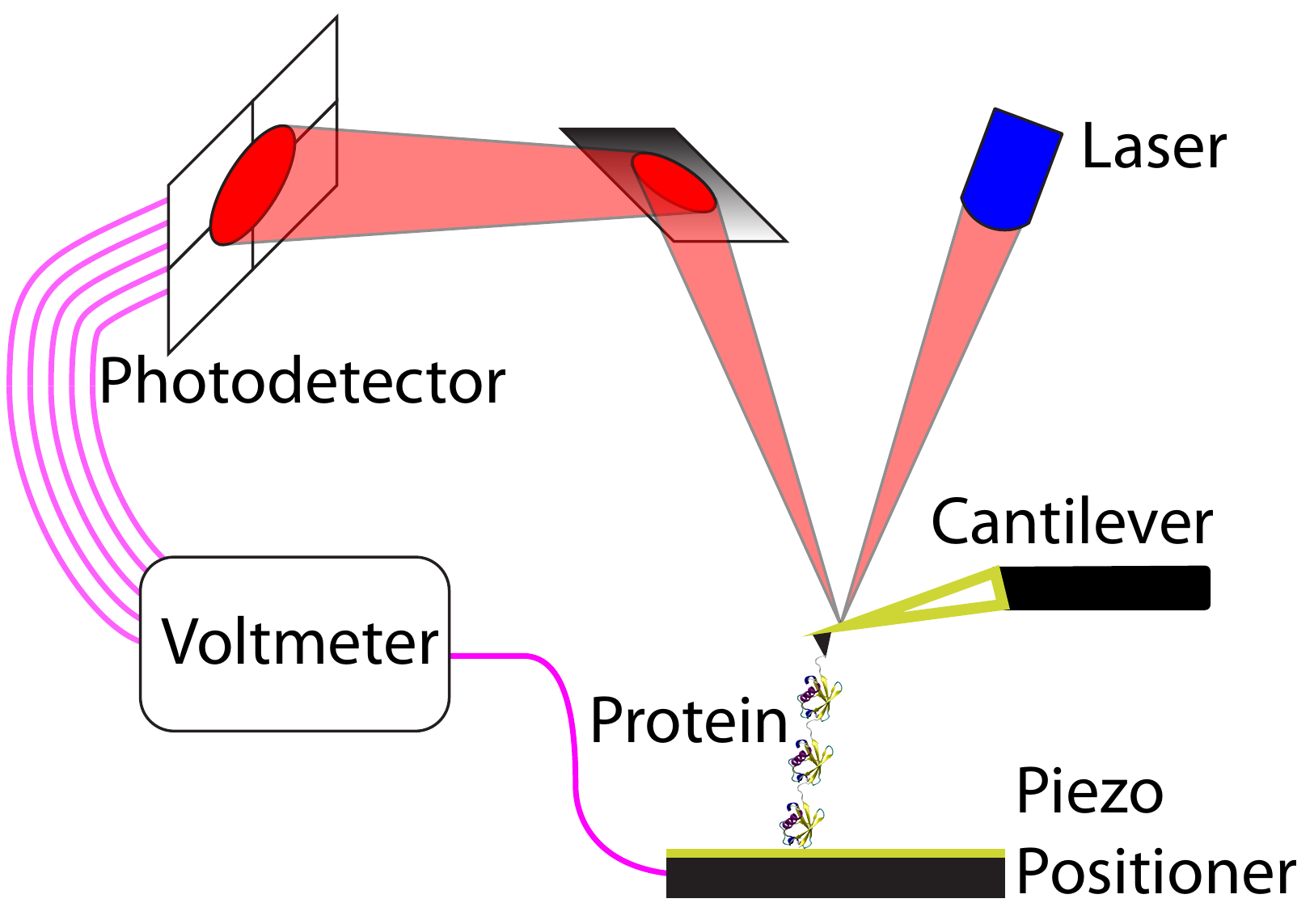}
\caption{\label{ExpSchem} (Color online) Schematic representation of a typical AFM setup for protein measurements. The deflection of the cantilever is measured as a differential voltage on the quadrant photodetector.  In order to achieve a force measurement one must convert the voltage measured on the photodector into a displacement of the tip and then measure the spring constant of the cantilever (typically done using the thermal tune method \cite{hutter1993}).}
\end{figure}
Non-exponential lifetime distributions in physical systems are often a signature of underlying complex processes with multiple timescales operating at the microscopic level \cite{palmer1984,weron1993,jurlewicz1993}.  Characterizing the range of rates present can provide insights on the fundamental physical mechanisms driving the decay process.  However, non-exponential distributions can also arise in systems that are characterized by a single timescale, if external fluctuations are able to propagate through non-linearities of the system.  Thus, if non-exponential behavior is to be taken as proof of complex internal workings the effects of external fluctuations must be understood.

\section{Random Bell model}

In his model for cellular adhesion \cite{bell1978}, G.I. Bell needed a relationship between rate and mechanical force.  He used a phenomenological relation that was found by Zhurkov and Kuksenko in a study of crack propagation in stressed bulk polymers \cite{zhurkov1975}.  In analogy with the Arrhenius relation at a given temperature $T$, this relation postulates an exponential relationship between the unfolding rate $\alpha$ and the applied external force $F$ given by
\bb \label{BellModel}
\alpha = \alpha_0 e^{\beta F x^\ddagger},
\ee
where $\alpha_0$ is the unfolding rate at zero applied force, $\beta = 1/k_B T$  and $x^\ddagger$ is a model parameter with units of length, usually interpreted as the distance to the transition state \cite{evans1997, izrailev1997, west2006}.  This relationship was subsequently referred to as the ``Bell model'' and applied to the dynamic strength of molecular bonds \cite{evans1997} and molecular measurements of protein unfolding by AFMs \cite{rief1997}.  This emperical relation is an approximation of the mean escape time from a well (Kramers' problem).  The general solution to such a problem depends on the details of the energy landscape and need not take a simple exponential form.  Deviations from this approximation will manifest as a force dependent prefactor $\alpha_0(f)$ and a force dependent $x^\ddagger(f)$ even in the absence of a complex or rough energy landscape \cite{dudko2006}.

In the application of this model to protein measurements configurational disorder enters as a source of fluctuation for $x^\ddagger$ and is sufficient to explain experimental results \cite{raible2006,kuo2010}. In principle, however, one could consider the effects on the lifetime distribution of fluctuations in all of the parameters, $\alpha_0, \beta, F$ and $x^\ddagger$. 

Fluctuations in $\alpha_0$ or $x^\ddagger$ would provide a window into the physics on the microscopic scale \cite{raible2006,kuo2010}, whereas fluctuations in $\beta$ and $F$ simply arise from the experimental conditions and provide no insight into protein physics.

We introduce the terminology \textit{Random Bell Model} to indicate a model similar to the Bell Model (Eq.\ref{BellModel}) in which each parameter is assumed to be randomly distributed.  We will probe this model to determine the impact of various sources of fluctuations on the resulting distributions of rates and lifetimes.\\

\section{Sources of fluctuations}

Most experiments are done at room temperature, without strict thermal regulation, leading to fluctuations in temperature.  However, even in a poorly controlled environment, the upper bound for thermal fluctuations will be on the order of 5\textdegree, leading to fluctuations in $\beta$ of less than 2\%.  For the sake of brevity we will not consider these fluctuations as being significant.  By definition, $\alpha_0$ is the rate at temperature $T$ and zero applied force, and should be inherent to the folded protein structure.  We will therefore assume that $\alpha_0$ does not fluctuate within a given protein, nor from one molecule to the next.  Note that if it proved relevant, it would be straightforward to extend the following results to include fluctuations in both $\beta$ and $\alpha_0$.

This leaves us with sources of fluctuations in $F$ and $x^\ddagger$ to consider.  As previously mentioned, fluctuations in $x^\ddagger$ are the physical quantity of interest which may be masked by fluctations in $F$.  In current experiments many sources of fluctuations in force are present.  Readily visible are the fluctuations in the force value observed over a single experimental trace.  These fluctuations include 
\begin{enumerate}
  \item Thermally induced fluctuations \cite{gittes1998} in the position and curvature of the cantilever tip as well as the position of the sample.
  \item Fluctuations induced at high frequencies by the force feedback system (setup dependent).
  \item Mechanical vibrations transmitted through the AFM to the cantilever tip or the sample (setup dependent) \cite{binnig1986}.
\end{enumerate}
The combined magnitude of these fluctuations has been estimated to be $\simeq 10\%$ for the apparatus in \cite{brujic2007}.  Less apparent, but likely more significant, are the fluctuations from experimental run to run induced by the calibration of the system (see Figure \ref{ExpSchem}).  These fluctuations include 
\begin{enumerate}
  \item Errors in the calibration of the spring constant of the cantilever (estimates vary from $10\%$ \cite{matei2006} to $20\%$ \cite{emerson2006} for the generally used thermal tune method of calibration).
  \item Time dependent drift in the spring constant of the AFM cantilever (on the order of 14\% in 60 minutes \cite{emerson2006}).
\end{enumerate}
  Therefore, a worst case estimate of the overall magnitude of the force fluctations would be between $20\%$ and $25\%$. 


%

%
\begin{figure}[t]
\includegraphics[width=.97 \columnwidth]{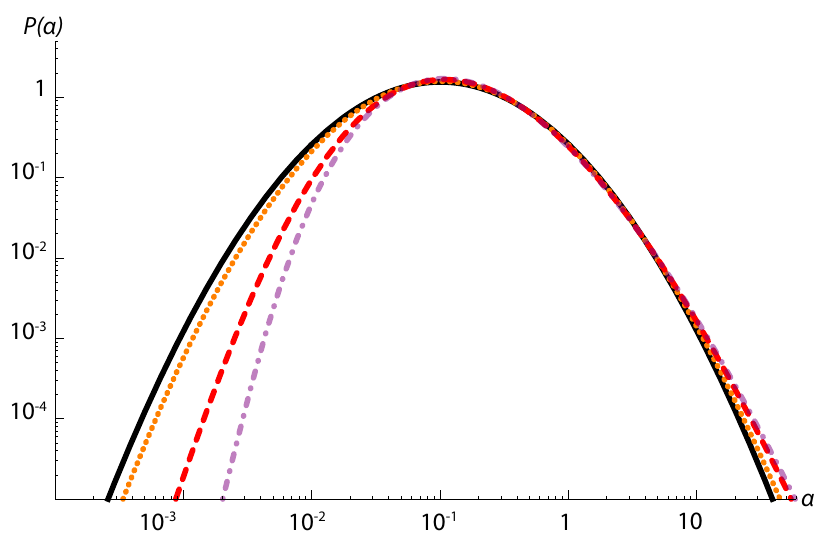}
\caption{\label{FigRate}(Color online) Probability density of rate $P(\alpha)$, given by Eq. \ref{RatePDF}, presented as a Log-Log plot. The curves correspond to values of $\beta \bar F \bar x^\ddagger = 6.11$ and $\alpha_0 = 10^{-3}\mathrm{s}^{-1}$, as given in reference \cite{kuo2010}, for a fixed value of $\sigma_\omega = 0.2$, but varying values of $(\sigma_\phi, \sigma_\xi) = (0,0.2)$ in black solid line, $(0.05, 0.19)$ in orange dotted line,$(0.1, 0.17)$ in red dashed line, $(0.15, 0.13)$ in purple dash dot line. Note that $(\sigma_\phi, \sigma_\xi) = (0,0.2)$ leads to a log-normal distribution of unfolding rates, represented by a parabola in Log-Log scales. }
\end{figure}

\section{Unfolding rate and time distributions}
\subsection{Unfolding rate distribution}
Let $F$ and $x^\ddagger$ be random variables with respective means $\bar F$ and $\bar x^\ddagger$ and standard deviations $\sigma_F$ and $\sigma_{x^\ddagger}$.  We introduce dimensionless random variables $\phi$ and $\xi$ such that
\begin{align}
F &= \phi \bar F ,\\
x^\ddagger &= \xi \bar x^\ddagger.
\end{align}
These two random variables are described by the joint probability density $J(\phi,\xi)$. 
By definition the probability density of the product $\omega=\phi \xi$ is given by 
\bb \label{defPw}
P_\omega(\omega) = \int_{-\infty}^{+\infty} d \phi d \xi  J(\phi,\xi) \delta(\omega - \phi \xi).
\ee
To simplify the previous relation, let's perform the following change of variables,
\bb
u_{1}=\phi,\\
u_{2}=\phi \xi,
\ee
whose Jacobian is 
\bb
\mathfrak{J}=\begin{vmatrix} 1 & 0  \\ \xi & \phi \end{vmatrix}=|\phi |=|u_{1}|.
\ee
Then Eq.\ref{defPw} becomes
\bb 
P_\omega(\omega) = \int_{-\infty}^{+\infty} d u_{1} d u_{2}  \frac{1}{|u_{1}|}J\left(u_{1},\frac{u_{2}}{u_1}\right) \delta(\omega - u_{2}).
\ee
The distribution of their product $\omega=\phi \xi$ is then given by the Rohatgi formula \cite{rohatgi1976}
\bb
\label{genProdPDF}P_\omega(\omega) &=& \int_{-\infty}^{+\infty}  \frac{d\xi}{|\xi|}J\left(\xi,\frac{\omega}{\xi}\right).
\ee
This expression of $P_{\omega}$ does not require $\phi$ and $\xi$ to be independent, nor of any particular distribution. To proceed further, however, we will make the experimentally reasonable assumption that the variables $\xi$ and $\phi$ are two independent Gaussian variables, with mean   1 by construction and standard deviations $\sigma_\xi = \sigma_{x^\ddagger}/\bar x^\ddagger$ and $\sigma_\phi = \sigma_F / \bar F$.  Note that we do not make any assumption on the relative importance of these two kinds of fluctuations. The distribution of $\omega$ is obtained by numerical integration of Eq.\ref{genProdPDF}. 

The distribution of rates $\alpha$ is obtained through the change of variables
\bb
\label{RandomBellModel} \alpha=\alpha_0 \exp \left( \beta \bar F \bar x^\ddagger \omega \right),
\ee
leading to the following probability density of unfolding rates :
\bb
\label{RatePDF} P_\alpha(\alpha)=\frac{1}{\beta \bar F \bar x^\ddagger \alpha} P_\omega \left[\frac{1}{\beta \bar F \bar x^\ddagger} \ln \left(\frac{\alpha}{\alpha_0} \right) \right].
\ee

\begin{figure}
\includegraphics[width=.97 \columnwidth]{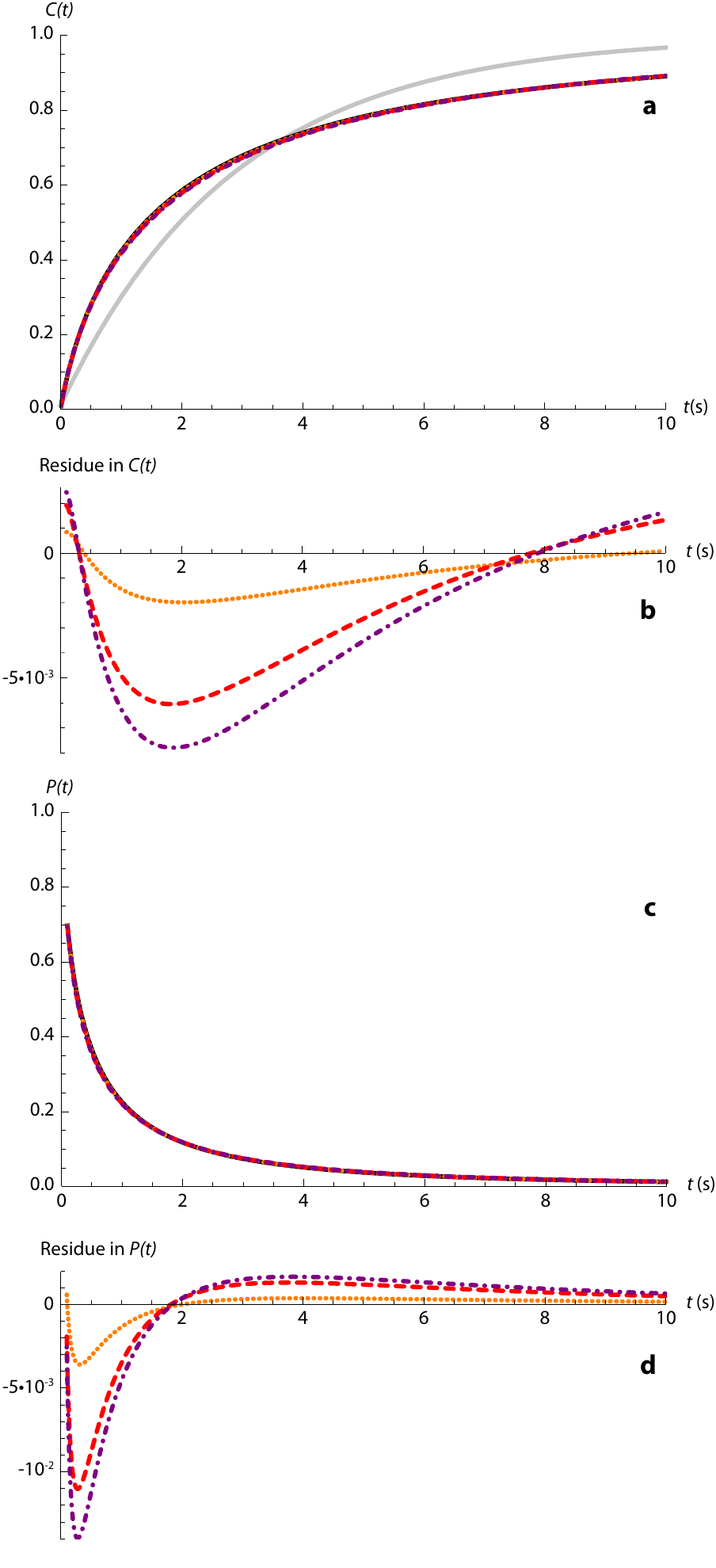}
\caption{\label{figPofTCofT}(Color online) Dependence of unfolding time statistics on the relative weight of external fluctuations ($\sigma_\phi$) and internal fluctuations ($\sigma_\xi$) for a fixed value of total fluctuations ($\sigma_\omega = 0.20$), and for $\beta \bar F \bar x^\ddagger = 6.11, \alpha_0 = 10^{-3}\mathrm{s}^{-1}$.  All four plots are for varying values of $(\sigma_\phi, \sigma_\xi) = (0,0.2)$ in black solid line, $(0.05, 0.19)$ in orange dotted line, $(0.1, 0.17)$ in red dashed line, $(0.15, 0.13)$ in purple dash dot line. a) Cumulative probability function of unfolding times. The grey line is the best exponential fit.  b) Difference between each curve in (a) and the curve for $(\sigma_\phi, \sigma_\xi) = (0, \sigma_\omega)$.  c) Probability density function of unfolding times.  d) Difference between each curve in (c) and the curve for $(\sigma_\phi, \sigma_\xi) = (0, \sigma_\omega)$.  The distributions shown in (a) and (c) are experimentally indistinguishable.}
\end{figure}

An example of such a rate distribution is presented in figure \ref{FigRate}.  We observe a systematic change in the tails of $P(\alpha)$ as the relative weight of $\sigma_\phi$ and $\sigma_\xi$ is varied.  As these changes are in the tails they may prove difficult to access experimentally.\\

\subsection{Unfolding time probability distributions}

The directly measurable cumulative unfolding time distribution is given by a Laplace transform of the rate distribution as
\bb \label{CumulTimes}
C(t)=1-\int_0^\infty d\alpha P_\alpha(\alpha) e^{-\alpha t}.
\ee
The corresponding probability density is then given by $P(t)=\frac{dC}{dt}(t)$.

Figures \ref{figPofTCofT}a and \ref{figPofTCofT}c show plots of the cumulative probability distributions and the probability density functions for a fixed value of $\sigma_\omega$, the standard deviation of the product variable $\omega$, and different values of $\sigma_\phi$ and $\sigma_\xi$.  The overall shape of the distribution appears to depend only on $\sigma_\omega$: The resulting distributions differ by less than 0.1\% and thus are indistinguishable for all values of $\sigma_\phi$ and $\sigma_\xi$ for a fixed value of $\sigma_\omega$.

If $\sigma_\phi$ very nearly equals zero (\textit{i.e.} no fluctuations in the externally applied force) no confusion is possible: any deviation from a strictly single exponential lifetime distribution must be attributed to internal fluctuations as described by $\sigma_\xi$  \cite{kuo2010}.  However this cannot be the case experimentally.
Given these sources of fluctuation it is unlikely that $\sigma_\phi$ is negligible. Consequently, a good estimate of $\sigma_\phi$ must be obtained before one can hope to learn anything about $\sigma_\xi$ or its origins.\\

\subsection{Skewness of the rate distribution}

We have shown that unfolding time statistics are largely insensitive to the origin of the fluctuations.  However, we note that the underlying rate distribution becomes asymmetric as the ratio between $\sigma_\phi$ and $\sigma_\xi$ grows, as shown in Figure \ref{FigRate}.  One way to separate the sources of disorder would be experimentally measure the assymetry of the rate distribution as characterized by the skewness.  At present such a measurement is experimentally inaccessible because unfolding rates are not experimentally measurable quantities.  Instead, they must be estimated from the observed unfolding times using statistical techniques with all of their incumbent limitations, drawbacks, and errors \cite{brujic2006,brujic2007}.  We include the following study of the skewness for two reasons.  First, it is the dominant evidence of the separability of internal and external fluctuations.  Second, we do so as a way to motivate the development of experimental tools to allow such a measurement to be effected.

The skewness $\gamma$ of the distribution $P(\alpha)$ is defined as
\bb \label{defskewness}
\gamma=\frac{\langle (\alpha-\bar \alpha)^3 \rangle}{\langle (\alpha-\bar \alpha)^2 \rangle^{3/2}},
\ee
where $\bar \alpha$ is the mean unfolding rate and $\langle \cdots \rangle$ stands for the average over $P_{\alpha}$.
\begin{figure}[bt]
\includegraphics[width=.97 \columnwidth]{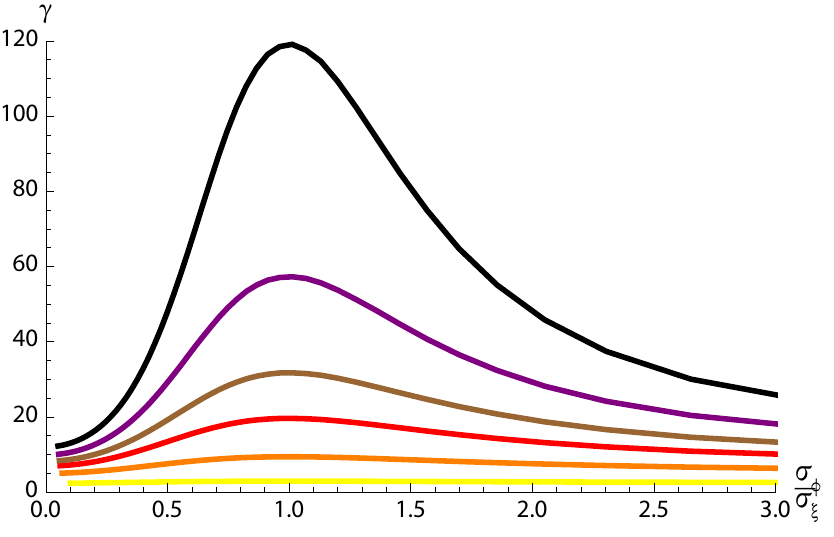}
\caption{\label{figSkewTheoretical}(Color online) Evolution of the rate distribution skewness $\gamma$ as a function of $\sigma_\phi/\sigma_\xi$, for 6 different values of $\sigma_\omega$, colored from top to bottom as $\sigma_\omega = 0.20$ (black), $0.19$ (purple), $0.18$ (brown) $0.17$ (red), $0.15$ (orange), and $0.10$ (yellow) for $\beta \bar F \bar x^\ddagger = 6.11,\ \alpha_0 = 10^{-3} \mathrm{s}^{-1}$.}
\end{figure}
\begin{figure}[bt]
\includegraphics[width=.97 \columnwidth]{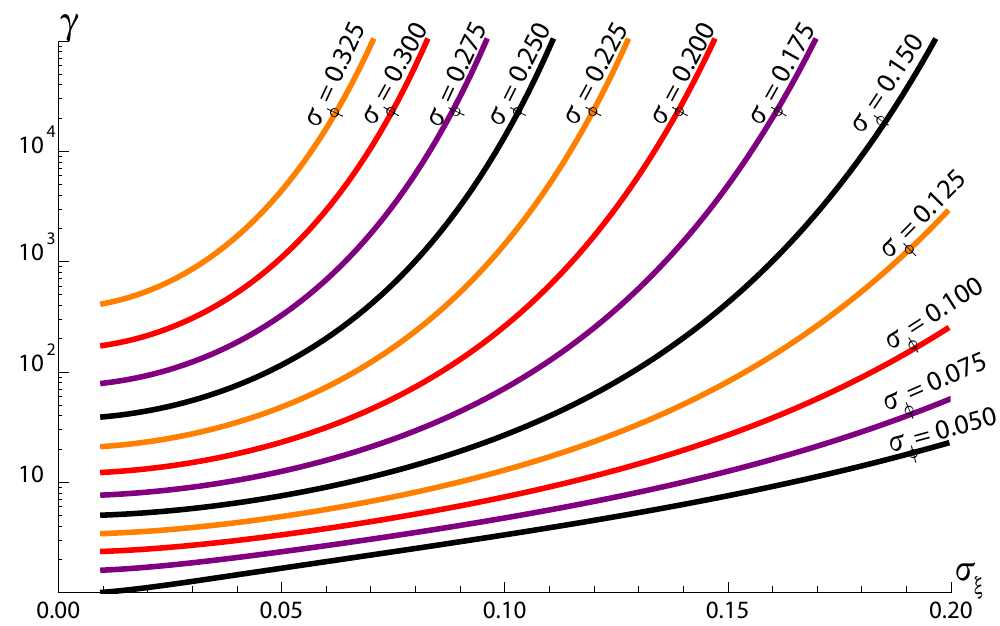}
\caption{\label{figSkewFixedSigPhi}(Color online) Evolution of the rate distribution skewness $\gamma(\sigma_{\phi},\sigma_{\xi})$ as a function of $\sigma_\xi$, for different values of $\sigma_\phi$, increasing by step of $0.025$ from $0.05$ (bottom black curve) to $0.325$ (top purple curve), with $\beta \bar F \bar x^\ddagger = 6.11,\ \alpha_0 = 10^{-3} \mathrm{s}^{-1}$.}
\end{figure}
Figure \ref{figSkewTheoretical} presents the evolution of the skewness $\gamma$ by varying the ratio of $\sigma_\phi$ and $\sigma_\xi$ for several fixed values of $\sigma_\omega$, the standard deviation of the product $\omega =\phi \xi$. We observe that the overall scale of the skewness grows with increasing $\sigma_\omega$ while the maximum, for fixed $\sigma_\omega$, always occurs when $\sigma_\phi = \sigma_\xi$. Note that skewness is of necessity symmetric under interchange of $\sigma_\phi$ and $\sigma_\xi$ because of the symmetry of the original problem.  Therefore, $\gamma$ must have a turning point when $\sigma_\phi = \sigma_\xi$.

Ideally, one should first accurately estimate $\sigma_\phi$, then a measurement of $\gamma$ will allow for $\sigma_\xi$ to be read off the graph presented in Figure \ref{figSkewFixedSigPhi}.  Perhaps more realistically, one can start with a fairly poor estimate of $\sigma_\phi$ and then measure $\gamma$ in a series of experiments in which additional, controlled, amounts of force fluctuation are injected into the experiment to allow for a differential measurement of $\sigma_\xi$.

\section{Conclusions}
In this article we analyzed the effects of both internal and external fluctuations on unfolding time statistics in the context of the random Bell model. We highlighted the extremely weak effect these different origins have on the measurable unfolding time distributions, as illustrated in Fig.\ref{figPofTCofT}. Thus it is difficult to attribute non-exponential kinetics solely to the physics of a protein. This observation ultimately comes from the fact that lifetime distribution measurements rely on the assumption that the unfolding times are independent and identically distributed (IID).  However, in the presence of external fluctuations this assumption may no longer be justified, particularly if data is gathered over the course of multiple days or weeks using multiple AFM cantilevers with multiple calibrations.  Even if the underlying behavior of a protein were perfectly exponential such a lumping together of the data will yield non-exponential lifetime distributions.  Statistical analyses, such as the maximum likelihood estimator \cite{aioanei2009}, might provide a way to relax the assumption of IID, at the cost of some arbitrariness in the choice of the likelihood function \cite{brujic2010}.

In this article we have focused on the application to constant force protein measurements, however, these arguments can be equally well applied to other afm experiments.  As an example, fluctuations in the calibration of experimental setups can have significant consequences on landscape reconstructions using force extension measurements \cite{imparato2008}.

Taken as a whole, these results indicate that it is extremely difficult to experimentally distinguish between internal and external sources of fluctuations by measuring lifetime distributions.  External fluctuations can mask the more interesting internal fluctuations.  As a necessary but not sufficient prerequisite to measuring the magnitude of internal fluctuations one must first minimize and quantify the level of external fluctuations.  Therefore, the internal physics will remain obscured unless AFM force experiments are accompanied by an estimation of the types and magnitudes of external fluctuations and the way that such fluctuations propogate through the measured quantities.

In light of these findings there is a need for new experimental methods in order to gain insight into the physics at play on the molecular scale without relying on a statistical analysis.  One promising route is to avoid statistical analysis of collected unfolding time by focusing on protein dynamics. In the spirit of nano-rheology experiments \cite{katz2009,wong2010}, direct measurements of the dynamical global response of individual molecules appears as a promising route to understand how interatomic interactions contribute to the mechanical properties of a protein.

\acknowledgements{ We would like to thank Dominique Bicout, Jasna Bruji\'c, Jean-Yves Fortin, Efim Kats, Herbert Lannon, John Royer and Timothy Ziman for discussions and comments.}


\end{document}